\input{aipcheck}

\documentclass[
    ,final            % use final for the camera ready runs
  ]
  {aipproc}

\layoutstyle{6x9}

\begin{document}

\title{Nuclear Star Clusters \& Black Holes}


\author{Anil Seth$^1$, Michele Cappellari$^2$, Nadine Neumayer$^3$, Nelson Caldwell$^1$, Nate Bastian$^4$, Knut Olsen$^5$, Robert Blum$^5$, Victor Debattista$^6$, Richard McDermid$^7$, Thomas Puzia$^8$, Andrew Stephens$^7$}{
  address={$^1$Harvard-Smithsonian Center for Astrophysics, Cambridge, USA, $^2$University of Oxford, UK, $^3$ESO, Garching, Germany, $^4$University of Cambridge, UK, $^5$National Optical Astronomy Observatory, $^6$Jeremiah Horrocks Institute, U. Central Lancashire, UK, $^7$Gemini, Hilo, USA, $^8$Herzberg Institute of Astrophysics, Victoria, Canada}
}

\begin{abstract}
We summarize the recent results of our survey of the nearest nuclear
star clusters.  The purpose of the survey is to understand nuclear
star cluster formation mechanisms and constrain the presence of black
holes using adaptive optics assisted integral field spectroscopy,
optical spectroscopy, and HST imaging in 13 galaxies within 5~Mpc.  We
discuss the formation history of the nuclear star cluster and possible
detection of an intermediate mass BH in NGC~404, the nearest
S0 galaxy.
\end{abstract}

\maketitle

%%%%%%%%%%%%%%%%%%%%%%%%%%%%%%%%%%%%%%%%%%%%
%% MAINMATTER
%%%%%%%%%%%%%%%%%%%%%%%%%%%%%%%%%%%%%%%%%%%%

\section{Introduction}

Nuclear star clusters (NSCs) are found at the centers of a majority of
galaxies with Milky Way mass and below
\citep{boker02,carollo02,cote06}.  These NSCs have similar sizes to
globular clusters ($\sim$5~pc), but are much more luminous and massive
\citep{walcher05}. Their morphology is often complicated and their star
formation history is typically extended \citep[e.g.][]{seth06}.  The
relationship of NSCs to supermassive black holes (BHs) typically found
at the centers of more luminous galaxies is not clear.  NSCs and
supermassive BHs co-exist in some galaxies \citep{seth08}, and both
have masses that scale with the mass and dispersion of the galaxy they
live in \citep[e.g.][]{ferrarese06}. Unlike BHs, the histories of
NSCs are recorded in their morphology, kinematics, and stellar
populations, making them excellent tools to study how
mass accumulates at the centers of galaxies.

\section{Nearby Nuclear Star Cluster Survey}

We are conducting an ongoing survey of the nearest NSCs, observing a
sample of 13 galaxies within 5~Mpc.  By resolving the kinematics and
stellar populations in these NSCs, we can understand how they
formed and constrain the masses of any BHs within them.  Their
proximity makes them among the most promising objects for detecting
intermediate mass BHs, constraining the low mass end of the BH-galaxy
scaling relations and the occupation fraction of low mass galaxies.
Because the BHs in these galaxies are likely to not have undergone
much subsequent evolution, measuring their properties is one of the
only ways of constraining the initial formation mechanism of
supermassive BHs
\citep[e.g.][]{volonteri08}.

We are obtaining adaptive optics assisted integral field spectroscopy
on the entire sample using Gemini/NIFS (PI: Seth) and VLT/SINFONI (PI:
Neumayer).  In addition, we are obtaining low resolution optical
spectroscopy to examine the stellar populations in the nucleus and HST
imaging to constrain their morphology.  Our first published results
\citep{seth08b}, show a flattened multi-component NSC with strong
rotation in the edge-on spiral galaxy NGC~4244.  This NSC must have
formed from repeated accretion of material from the disk of the host
galaxy, a scenario tested in simulations by Markus Hartmann (this
volume).

\begin{figure}
  \includegraphics[height=3in]{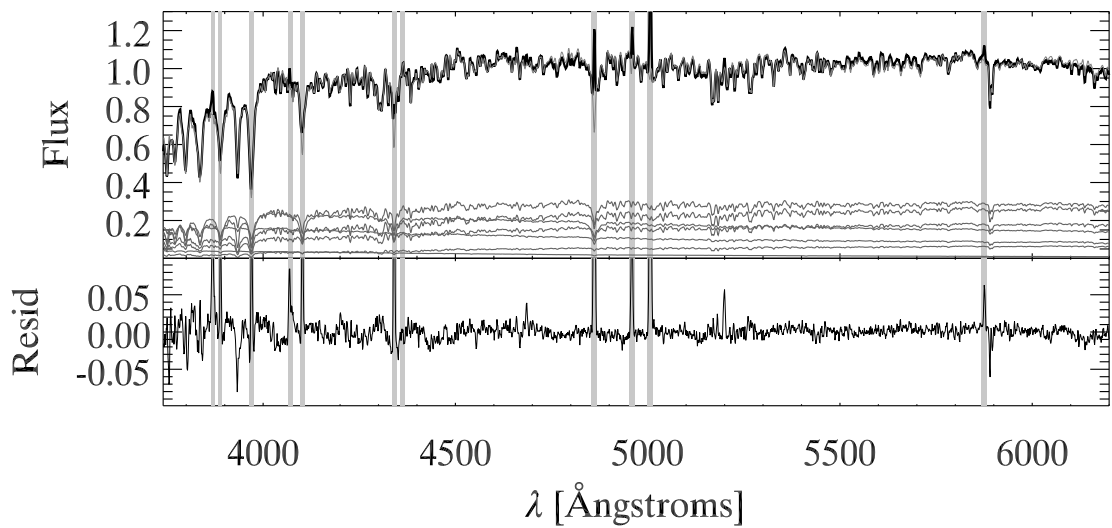}

 \caption{MMT 6.5m optical spectrum of the NGC~404 nuclear cluster
(black line) with the best fit spectral synthesis model (overplotted
gray line).  The spectral templates used to fit the spectra (shown in
light gray) are drawn from 11 possible age bins ranging from 1~Myr to 13~Gyr
in age.  More than 50\% of the light is in stars with ages 1-3~Gyr,
while $\sim$10\% is at ages $\leq$10~Myr.  Residuals are about 1\%.
Vertical gray lines indicate areas excluded from the fit due to
emission lines.  \label{aseth_specfig}}
  
\end{figure}

\section{The Nucleus of NGC~404}

We now focus on the NSC formation and possible intermediate mass BH in
NGC~404, the nearest S0 galaxy (D=3.1~Mpc).  The galaxy is nearly
face-on and has a total mass of about $10^9$~M$_\odot$ and $M_V =
-17.35$.  NGC~404 also hosts the nearest LINER nucleus, although
whether this observed activity is actually the result of an accreting
BH is a matter of some contention.  The results shown here are
presented in much greater detail in a paper submitted to the {\em
ApJ}.

\vspace{0.03in}

\begin{figure}
  \includegraphics[height=2.5in]{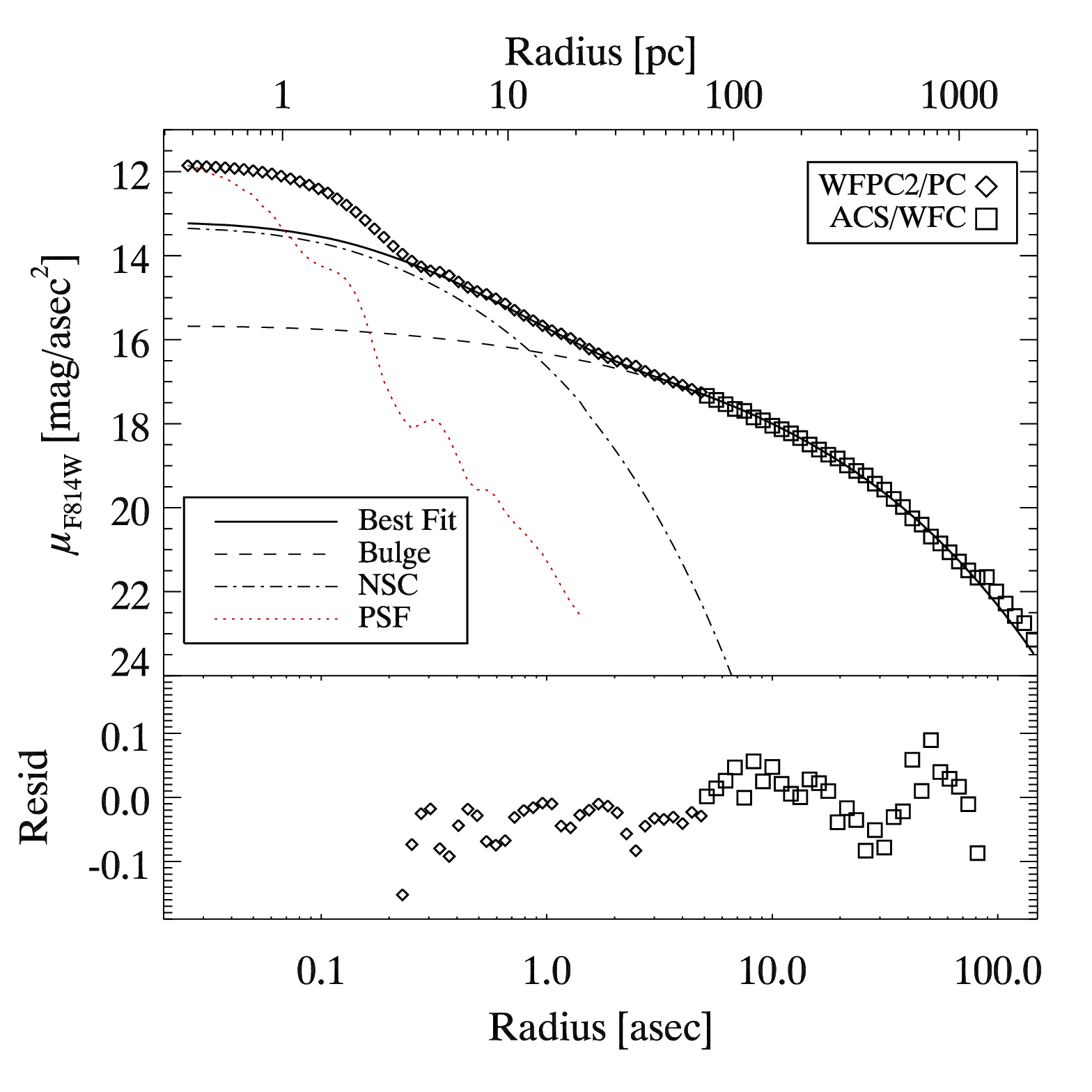}
    \includegraphics[height=2.5in]{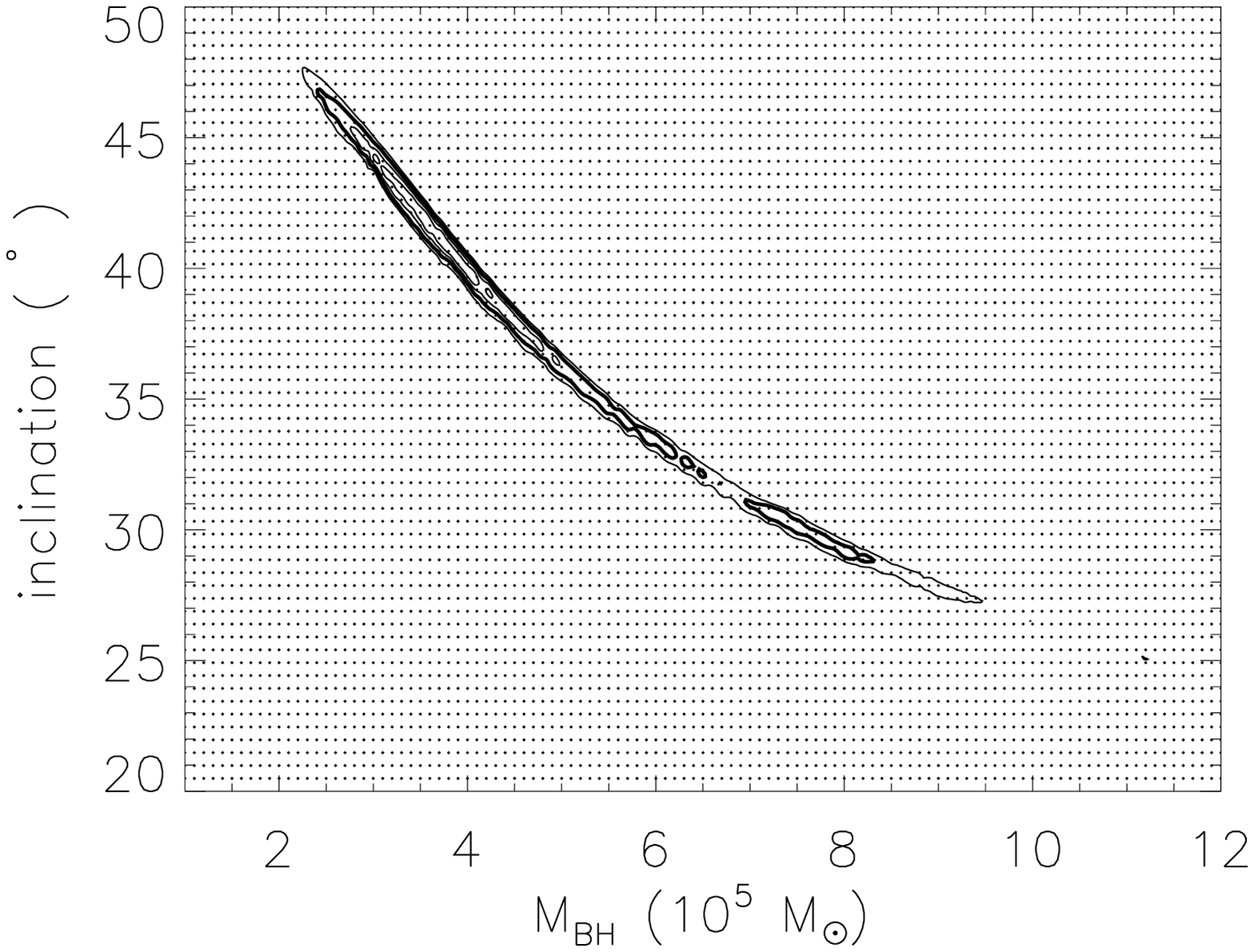}
    \caption{{\em Left --} The surface brightness profile of NGC~404
    in $F814W$/$I$ band made from HST imaging.  The lines show the
    best bulge$+$NSC fit, where both components are assumed fit as
    S\'ersic profiles. The central 0.2'' show a clear excess
    above these profiles. {\em Right} -- Results from the gas dynamical
    model of a disk of molecular H$_2$ gas emission.  Shown are
    $\chi^2$ contours as a function of BH mass and disk inclination.
    The best fitting BH mass is $4.5^{+3.5}_{-2.0} \times
    10^5$~M$_\odot$, however the stellar dynamical models suggest a BH
    mass upper limit of $1 \times 10^5$~M$_\odot$.
 \label{aseth_nscfig}}
  
\end{figure}

\noindent {\bf Nuclear Star Cluster}\\
The surface brightness profile of the inner part of NGC~404 (see left
panel of Fig.~\ref{aseth_nscfig}) can be decomposed into three
distinct components each with distinct kinematics and stellar
populations.  The bulge of the galaxy has a mass of $9 \times
10^8$~M$_\odot$ and dominates the light profile outside the central
arcsecond (15~pc).  Optical spectra of the stellar populations of this
component show that it is dominated by old stars with ages $>$5~Gyr.
The nuclear star cluster dominates the light in the central arcsecond
and has a half-light radius of 10~pc and a dynamical mass of
$10^7$~M$_\odot$, larger than expected from NSC scaling relations.  Its
optical spectrum is dominated by the light of stars with age of
1-3~Gyr, with stellar population synthesis fits suggesting that over
half the mass of the cluster was formed during this period.  The
cluster shows mild rotation in the same direction as HI gas at larger
radii in the galaxy \citep{delrio04}, which has been suggested to have
a merger origin.  Finally, within the central 0.2'', the central
excess of emission seen in the surface-brigthtness profile, is also
found to be counter-rotating relative to the NSC.  The emission of
this component probably results from a combination of young
($<$100~Myr) stars, hot dust emission and perhaps AGN continuum.

The nuclear star cluster in NGC~404 appears to have a quite different
formation from the episodic disk accretion seen in NGC~4244.  The
burst of star formation $\sim$1~Gyr ago is suggestive of a merger
origin, a scenario that is also consistent with the difference in
rotation between the central excess and nuclear star cluster.

\vspace{0.03in}

\noindent {\bf Possible Intermediate Mass BH}\\
Signs of BH accretion in NGC~404 are conflicting, with the
strongest evidence coming from the presence of variable UV emission
\citep{maoz05}.  We also find compact and possibly variable hot dust
emission indicative of BH accretion at the center of the NSC. 

From the Gemini/NIFS data we have obtained kinematic measurements of
both the stars (as measured by the CO-bandhead absorption at
2.3$\mu$m) and for a gas disk seen in emission of excited H$_2$ at
2.12$\mu$m.  The NIFS adaptive optics observations have a PSF core
with a FWHM=0.09'' containing 50\% of the light.  We have created
dynamical models to fit to the kinematics of both tracers.  The
starting point of both models is a mass model for the galaxy generated
from HST imaging data using the multi-gaussian expansion method
\citep{emsellem94}.  The mass model assumes a constant mass-to-light
($M/L$) ratio. 
%HST color maps suggest and the central excess suggest that the $M/L$
%may decline at the center, an effect which could decrease the mass of
%the BH by up to $3 \times 10^5$~$M_\odot$ in our constant $M/L$
%dynamical models.

The stellar dynamical model uses axisymmetric Jeans modeling with the
JAM package \citep[][see also contribution in this
volume]{cappellari08} to fit the CO bandhead velocity and dispersion
in the central 1.5'' (22~pc) of the galaxy. The model fits the NSC
$M/L$, the anisotropy ($\beta_z$), and the BH mass ($M_{\rm BH}$).
Assuming a constant $M/L$, we find a $M/L=0.70$ in the $I$ band and a
3-$\sigma$ upper limit of $M_{\rm BH} < 1 \times 10^5$~M$_\odot$.  The
$M/L$ determined from the dynamical model matches the best-fit stellar
population synthesis $M/L$ to within 10\%.

The gas dynamical model is similar to that presented in
\citet{neumayer07}, and assumes a thin disk geometry.  It uses the $M/L$ from
the stellar dynamical model and finds the disk inclination and BH mass
that best match the H$_2$ kinematics.  The results of this fit are
shown in the right panel of Fig.~\ref{aseth_nscfig}.  The best-fit
$M_{\rm BH} = 4.5^{+3.5}_{-2.0}
\times 10^5$~M$_\odot$ (3$\sigma$ errors), somewhat inconsistent with
the BH mass upper
limit from the stellar kinematics.  However both BH mass estimates are
comparable to the uncertainty in the central stellar mass of $< 3
\times 10^5$~M$_\odot$ due to possible $M/L$ variations within the
cluster.  Despite this uncertainty, the dynamical models constrain the
mass of the possible BH to be smaller than any previously dynamically
measured BH and below several bulge mass scaling relations.  We hope
to obtain HST spectroscopy and imaging of the stellar populations
across the nucleus to better refine the mass model and robustly
measure the BH mass in NGC~404.

%%%%%%%%%%%%%%%%%%%%%%%%%%%%%%%%%%%%%%%%%%%%
%% SAMPLE TABLE
%%
%% Shows the use of \tablehead and \tablenote
%% macros
%%%%%%%%%%%%%%%%%%%%%%%%%%%%%%%%%%%%%%%%%%%%

%% \begin{table}
%% \begin{tabular}{lrrrr}
%% \hline
%%   & \tablehead{1}{r}{b}{Single\\outlet}
%%   & \tablehead{1}{r}{b}{Small\tablenote{2-9 retail outlets}\\multiple}
%%   & \tablehead{1}{r}{b}{Large\\multiple}
%%   & \tablehead{1}{r}{b}{Total}   \\
%% \hline
%% 1982 & 98 & 129 & 620    & 847\\
%% 1987 & 138 & 176 & 1000  & 1314\\
%% 1991 & 173 & 248 & 1230  & 1651\\
%% 1998\tablenote{predicted} & 200 & 300 & 1500  & 2000\\
%% \hline
%% \end{tabular}
%% \caption{Average turnover per shop: by type
%%   of retail organisation}
%% \label{tab:a}
%% \end{table}

%\begin{theacknowledgments}
%Acknowledgments
%\end{theacknowledgments}



\bibliographystyle{aipproc}   % if natbib is available

%\bibliography{/Users/seth/allreferences}

\begin{thebibliography}{14}
\expandafter\ifx\csname natexlab\endcsname\relax\def\natexlab#1{#1}\fi
\providecommand{\enquote}[1]{``#1''}
\expandafter\ifx\csname url\endcsname\relax
  \def\url#1{\texttt{#1}}\fi
\expandafter\ifx\csname urlprefix\endcsname\relax\def\urlprefix{URL }\fi
\providecommand{\eprint}[2][]{\url{#2}}

\bibitem[{B{\"o}ker} et~al.(2002)]{boker02}
T.~{B{\"o}ker}, S.~{Laine}, R.~P. {van der Marel}, M.~{Sarzi}, H.-W. {Rix},
  L.~C. {Ho}, and J.~C. {Shields}, \emph{AJ} \textbf{123}, 1389--1410 (2002).

\bibitem[{Carollo} et~al.(2002)]{carollo02}
C.~M. {Carollo}, M.~{Stiavelli}, M.~{Seigar}, P.~T. {de Zeeuw}, and
  H.~{Dejonghe}, \emph{AJ} \textbf{123}, 159--183 (2002).

\bibitem[{C{\^o}t{\'e}} et~al.(2006)]{cote06}
P.~{C{\^o}t{\'e}}, S.~{Piatek}, L.~{Ferrarese}, A.~{Jord{\'a}n}, D.~{Merritt},
  E.~W. {Peng}, M.~{Ha{\c s}egan}, J.~P. {Blakeslee}, S.~{Mei}, M.~J. {West},
  M.~{Milosavljevi{\'c}}, and J.~L. {Tonry}, \emph{ApJS} \textbf{165}, 57--94
  (2006).

\bibitem[{Walcher} et~al.(2005)]{walcher05}
C.~J. {Walcher}, R.~P. {van der Marel}, D.~{McLaughlin}, H.-W. {Rix},
  T.~{B{\"o}ker}, N.~{H{\"a}ring}, L.~C. {Ho}, M.~{Sarzi}, and J.~C. {Shields},
  \emph{ApJ} \textbf{618}, 237--246 (2005).

\bibitem[{Seth} et~al.(2006)]{seth06}
A.~C. {Seth}, J.~J. {Dalcanton}, P.~W. {Hodge}, and V.~P. {Debattista},
  \emph{AJ} \textbf{132}, 2539--2555 (2006).

\bibitem[{Seth} et~al.(2008{\natexlab{a}})]{seth08}
A.~C. {Seth}, M.~{Ag{\"u}eros}, D.~{Lee}, and A.~{Basu-Zych}, \emph{ApJ}
  \textbf{678}, 116--130 (2008{\natexlab{a}}).

\bibitem[{Ferrarese} et~al.(2006)]{ferrarese06}
L.~{Ferrarese}, P.~{C{\^o}t{\'e}}, E.~{Dalla Bont{\`a}}, E.~W. {Peng},
  D.~{Merritt}, A.~{Jord{\'a}n}, J.~P. {Blakeslee}, M.~{Ha{\c s}egan},
  S.~{Mei}, S.~{Piatek}, J.~L. {Tonry}, and M.~J. {West}, \emph{ApJL}
  \textbf{644}, L21--L24 (2006).

\bibitem[{Volonteri} et~al.(2008)]{volonteri08}
M.~{Volonteri}, G.~{Lodato}, and P.~{Natarajan}, \emph{MNRAS} \textbf{383},
  1079--1088 (2008).

\bibitem[{Seth} et~al.(2008{\natexlab{b}})]{seth08b}
A.~C. {Seth}, R.~D. {Blum}, N.~{Bastian}, N.~{Caldwell}, V.~P. {Debattista},
  and T.~H. {Puzia}, \emph{ApJ} \textbf{687}, 997--1003 (2008{\natexlab{b}}).

\bibitem[{del R{\'{\i}}o} et~al.(2004)]{delrio04}
M.~S. {del R{\'{\i}}o}, E.~{Brinks}, and J.~{Cepa}, \emph{AJ} \textbf{128},
  89--102 (2004).

\bibitem[{Maoz} et~al.(2005)]{maoz05}
D.~{Maoz}, N.~M. {Nagar}, H.~{Falcke}, and A.~S. {Wilson}, \emph{ApJ}
  \textbf{625}, 699--715 (2005).

\bibitem[{Emsellem} et~al.(1994)]{emsellem94}
E.~{Emsellem}, G.~{Monnet}, and R.~{Bacon}, \emph{A\&A} \textbf{285}, 723--738
  (1994).

\bibitem[{Cappellari}(2008)]{cappellari08}
M.~{Cappellari}, \emph{MNRAS} \textbf{390}, 71--86 (2008).

\bibitem[{Neumayer} et~al.(2007)]{neumayer07}
N.~{Neumayer}, M.~{Cappellari}, J.~{Reunanen}, H.-W. {Rix}, P.~P. {van der
  Werf}, P.~T. {de Zeeuw}, and R.~I. {Davies}, \emph{ApJ} \textbf{671},
  1329--1344 (2007).

\end{thebibliography}

\IfFileExists{\jobname.bbl}{}
 {\typeout{} \typeout{******************************************}
  \typeout{** Please run "bibtex \jobname" to optain} \typeout{** the
  bibliography and then re-run LaTeX} \typeout{** twice to fix the
  references!}  \typeout{******************************************}
  \typeout{} }

\end{document}